\documentstyle[preprint,aps,epsf]{revtex}

\begin{document}
\draft
\title{Experimental studies of Chaos and Localization in Quantum Wavefunctions}
\author{A.Kudrolli, V.Kidambi and S.Sridhar}
\address{Department of Physics, Northeastern University, Boston, MA 02115.}
\date{December 23rd, 1994}
\maketitle

\begin{abstract}
Wavefunctions in chaotic and disordered quantum billiards are studied
experimentally using thin microwave cavities. The chaotic wavefunctions
display universal density distributions and density auto-correlations in
agreement with expressions derived from a 0-D nonlinear $\sigma $-model of
supersymmetry, which coincides with Random Matrix Theory. In contrast,
disordered wavefunctions show deviations from this universal behavior due to
Anderson localization. A systematic behavior of the distribution function is
studied as a function of the localization length, and can be understood in
the framework of a 1-D version of the nonlinear $\sigma $-model.
\end{abstract}

Complexity in Quantum Mechanics can arise from two sources :\ chaos and
disorder. Non-Integrable systems which are classically chaotic are now known
to be of relevance to a variety of atomic and nuclear phenomena \cite
{gutzwiller90}. Disorder is of fundamental importance in condensed matter
physics, where it leads to the phenomenon of localization \cite{anderson58}.
For quantum chaotic systems, Random Matrix Theory (RMT) has been shown to
provide a very good description of universal statistical properties of
eigenvalue spectra \cite{bohigas84,mehta90}. Recent developments in the
theory of disordered systems based upon nonlinear $\sigma $-models using
supersymmetry theory \cite{efetov83,prigodin94a,mirlin93a} have led to the
recognition that the extreme diffusive limit of disordered systems also
behave similarly to quantum chaotic systems. These theories have made
several quantitative predictions for correlations in mesoscopic systems,
including the contribution due to localization. However these ideas remain
to be tested experimentally. Experiments also have added importance as
several systems of Quantum Chaos which have localization, may behave
similarly \cite{fyodorov94}, for e.g. the kicked rotator, which has been
studied numerically shows deviation from RMT behavior \cite{casati90}.

While the eigenvalue statistics of quantum chaotic systems has been
experimentally tested, the eigenfunction statistics has not been done
before. The principal reason is the lack of accessibility to wavefunctions.
In atoms and nuclei, the nature of the wavefunctions only manifests
indirectly in quantities such as transition rates. Experiments using
microwave cavities which exploit the correspondence between the Maxwell and
Schr\"odinger equations, are unique in that they allow direct measurement of
the eigenfunctions, as well as eigenvalues, in model billiard geometries.
Earlier such experiments have provided direct observation of scars \cite
{sridhar91}, enabled precise tests of eigenvalue statistics\cite{kudrolli94a}%
, and have demonstrated ability to study arbitrary geometries which are not
accessible to numerical simulation. In this paper we use such experiments to
study the influence of chaos and localization on quantum wavefunctions.

The experiments were carried out using thin $(d<6$ mm$)$ cavities, whose
cross-sections can be shaped in essentially arbitrary geometries. For these
two dimensional cavities, the operational wave equation is $(\nabla
^2\,+\,k^2)\Psi =0$. The details of the experimental method are described in
ref. \cite{sridhar92a}. Eigenfunctions were directly measured using a cavity
perturbation technique which measures $|\Psi |^2$ \cite{sridhar91,sridhar92a}%
. We study the manifestations of chaos and localization in terms of
statistical properties of the eigenfunctions, such as the density
probability distribution $P(|{\Psi }|^2)$ and the density auto-correlation $%
P_2(r)=\ <|{\Psi (\stackrel{\rightarrow }{q})}|^2|{\Psi (\stackrel{%
\rightarrow }{r}+\stackrel{\rightarrow }{q})}|^2>$, and the Inverse
Participation Ratio (IPR), $P_2(r\rightarrow 0)$.

The geometries studied were representative of integrable, chaotic and
disordered systems. Among chaotic systems, besides the quarter Sinai
billiard, we also study the Sinai-Stadium geometry introduced by us in ref. 
\cite{kudrolli94a}. The Sinai-Stadium has isolated periodic orbits
(PO), and gives direct and exact agreement with RMT for the eigenvalue
statistics, unlike the Sinai billiard and the stadium billiard for which the
nonisolated PO contribution leads to deviations from universality \cite
{kudrolli94a}. Representative eigenfunctions for some of the geometries are
shown in Fig.1. The 5.685 GHz (Fig. 1, top left) state of the Sinai-Stadium
is similar to over almost a hundred other states of different energies that
were observed, in that there are no {\it visible} scars that can be
evidently associated with a PO. Indeed scarred states are few and the 7.370
GHz state shown in Fig. 1 (top right) is a rare coincidence with a
whispering gallery PO.

Localization effects were observed by fabricating billiards in which tiles
were placed to act as hard scatterers. These geometries are obtained by
placing 1cm square or circular tiles in a 44 cm x 21.8 cm rectangle at
random locations (Fig. 1). (The locations were generated using a random
number generator and the tiles placed manually). Earlier experiments \cite
{sridhar93} on the eigenvalue spectrum of similar disordered geometries had
shown that these experiments exemplify textbook two dimensional electron
systems without interactions to remarkable degree. To our knowledge, these
geometries are difficult to study using numerical computation and hence the
microwave experiments at present afford the only reliable means to study
this type of disordered systems. More than one realization of the disordered
geometry was experimentally studied, labeled D1 through D5 depending on the
density of scatterers (Table 1).

Sample eigenfunctions of D3, are shown in Fig. 1. Here again we have
obtained nearly 100 wavefunctions. It is evident that in these billiards,
any association with classical structures such as periodic orbits is
difficult to see, although of course these billiards are also chaotic and
such association must exist in principle. Instead the most striking effect
visible is localization, which is strongest in the 3.372 GHz state, but is
also present in a weaker form in the 6.651 GHz state. The degree of
localization can be varied either by changing the frequency window for a
given geometry, or changing the density of tiles. Hence the mean free path $l
$ and also the localization length $\xi $ could be manipulated, the latter
over almost two orders of magnitude. 

We first carry out an analysis of the chaotic wavefunctions. $P(|{\Psi }|^2)$
is shown in Fig.2 for the chaotic geometries - the Sinai-Stadium and the
Sinai billiard. These are compared with the well-known Porter Thomas (P-T)
distribution obtained from RMT \cite{haake91}:

\begin{equation}
P(|{\Psi }|^2)=\frac 1{\sqrt{2\pi |{\Psi }|^2}}\exp (-\frac{|{\Psi }|^2}2) 
\end{equation}

The best agreement is shown by the Sinai-Stadium. The Sinai billiard
displays slight deviations which we attribute to states influenced by
bouncing ball orbits. In an integrable system, the probability distribution
is often truncated, e.g. for a rectangle at $|{\Psi }|^2=4$, and is not
universal. In contrast, the chaotic wavefunctions show a finite, although
exponentially vanishing probability of finding large intensities, in exact
agreement with the P-T distribution, thus confirming their universal nature.
Numerical simulations of random superpositions of plane 
waves was also done following ref. \cite{o'connor88} and also 
shows very good agreement with P-T \cite{berry81}.

The density correlation $P_2(r)$ was also determined from the wavefunctions,
and is shown in Fig.3. The angular brackets denote averaging over space $%
\stackrel{\rightarrow }{q}$ , over angles between $\stackrel{\rightarrow }{q}
$ and $\stackrel{\rightarrow }{r}$ , and finally over several states. An
important aspect of the present work is that our experimental results for
the chaotic and weak disorder cavities (see later) are well fit by the
functional form :

\begin{equation}
P_2(r)\;=1+c\,J_0^2(kr) 
\end{equation}

where $k$ is the wave number. When computed for individual states, the
result for $P_2(r)$ still has the functional form of Eq.(2), although the $%
r\rightarrow 0$ value is slightly different. However when averaged over
several states, the resultant $P_2(r)$ is a robust quantity determined by
the geometry alone. We further emphasize that we have checked and confirmed
that $P_2(r)$ is isotropic for the Sinai-Stadium.
The asymptotic value of $P_2(r\rightarrow \infty )=1$ can be understood
since $P_2(r\rightarrow \infty )\rightarrow <|{\Psi (\stackrel{\rightarrow }{%
q})} |^2><|{\Psi (\stackrel{\rightarrow }{q})}|^2>\rightarrow 1$ as the 
wavefunctions are completely decorrelated at larger distances. 

The data for the Sinai-Stadium yield $c=2\pm 0.1$. The Sinai billiard yields 
$c=1.7\pm 0.1$. An understanding of this value of $c=2$ can be achieved by
noting that in general, $P_2(r\rightarrow 0)=<|{\Psi (\stackrel{\rightarrow 
}{q})}|^4>=1+c$.  If $x=|{\Psi (\stackrel{\rightarrow }{q})}|^2$, then $%
P_2(r\rightarrow 0)=\int_0^\infty x^2P(x)dx$. Hence from Eq. 1 we get $%
P_2(r\rightarrow 0)=3$. Thus the value of $c=2$ for the chaotic case are
consistent with the P-T distribution in Fig.2.

The functional form in Eq.(2) with $c=2$ has been anticipated by Prigodin,
et. al. \cite{prigodin94a,prigodin94b} based upon a 0-D $\sigma $-model of
supersymmetry. This theory is expected to apply to disordered systems in the
diffusive limit, where the wavelength $\lambda <<l$ (scattering length) $<<L$%
, the cavity size. Results derived from this theory were earlier seen to
coincide with RMT \cite{efetov83}. Therefore it gives P-T distribution for
the eigenfunction components as in RMT. But it also gives coordinate
correlations not available from RMT alone \cite{prigodin94a,prigodin94b}. On
the basis that for the diffusive limit, the randomization due to disorder of
an infinite system is the same as that due to chaos at the boundary of a
finite system, the results should apply to chaotic systems. Our experiments
confirm both these predictions of the 0-D $\sigma$-model for $P(|{\Psi }|^2)$
and $P_2(r)$ in the chaotic cavities.

We now turn to an analysis of the eigenfunctions of the ``disordered''
geometries. We have divided the obtained wavefunctions into two regimes,(a) $%
\lambda >l$, i.e. the wavelength is greater than the mean free path. Here
one observes strongly localized regions (Fig.1, bottom left). (b) At higher
frequencies, where $\lambda <l$, the disordered eigenfunctions have the
expected feature of having maxima apparently randomly distributed over the
available domain (Fig.1, bottom right). In the experimental window from 2 to
8 GHz (15 to 3.2 cm), D1 is in regime (b), D2, D3 and D4 go from regime (a)
to (b), and D5 is in regime (a).

Fig.4 displays the density probability $P(|{\Psi }|^2)$ for D3 and D4, in
regime (b) for data in the 6.5 to 7.75 GHz range. The disordered data are
clearly not in agreement with the P-T distribution, displaying consistent
and significant deviations at the high intensity end which are well beyond
experimental error. These correspond to localized regions present in the
disordered wavefunctions, in which the peak intensity is large compared to
the average intensity.

Recently, Mirlin and Fyodorov have been able to calculate the localization
corrections in a quasi one dimensional wire using a 1-D nonlinear $\sigma $%
-model of supersymmetry \cite{mirlin93a}. They have been also able to extend
it to the full two dimensional case of the cavity in the delocalized regime 
\cite{fyodorov94c}. For incipient localization, the results can be expressed
as corrections to the P-T distribution given by:

\begin{equation}
P(x)=\,f(x)\exp (-x/2)/\sqrt{2\pi x}\,\, 
\end{equation}

where $x=|{\Psi (\stackrel{\rightarrow }{q})}|^2$ as before and $%
f(x)=1+d(1/4-x/2+x^2/12)$ for small $d$. Now d depends on localization
length $\xi $ and the density of disorder $l/L$ through $d=L/(2\pi \xi
)ln(L/l)$ for the two dimensional case \cite{fyodorov94c}. This form is
plotted in Fig. 3 for D3 and shows that the 1-D $\sigma $-model gives an
accurate description of the localization effects seen in the experiment.
From the fit to data for D1 to D4 in the 6.5 to 7.75 GHz range, $\xi /L$ are
extracted and listed in Table 1.

$P_2(r)$ for D3 is shown in Fig.5 and is well described by the functional
form Eq.2. The $c$ value obtained from a fit is given in
Table 1. This is consistent, within experimental error, with the values
obtained directly from the density distribution in Eq. 3, by using $%
IPR=\int_0^\infty x^2P(x)dx=1+c=3+d[L/(\xi \pi )]ln(L/l)$. This higher value
is therefore due to the high intensity peaks observed in Fig.1 and the
non-P-T distribution function in Fig.4 and depends on the localization
length.

In the strong localization regime (a) the deviations are even greater as the
effects due to localization is even more pronounced. For strong
localization, like in D5, stronger deviations in $P(|{\Psi }|^2)$ are seen
which cannot be described by corrections to P-T form as in Eq. 3. However
the data is described well by expressions for strong localization, for the
wire case, in ref. \cite{mirlin93a}. The form, $P(x)=8/(L/\xi )^2\sqrt{%
(L/\xi )/2x}K_1(2\sqrt{2x/(L/\xi )})$ fits well and a $\xi /L$ value of 0.1 
can be extracted for the data in the 3 to 6 GHz range, however some
constants due to dimensionality considerations are expected to change. $%
P_2(r)$ also does not obey the expression Eq. 2, as is evident from Fig.5.
Instead spatial correlations die out faster, perhaps exponentially. Although
the experimental data are available, further theoretical work needs to be
done to compare the data in the strong and intermediate localization regime.

An important perspective can be gained by examining the IPR for different
eigenstates that were measured for both the chaotic and the D3 and D4
cavities, as shown in the inset to Fig.5. It is seen to fluctuate from one
eigenstate to the other - such level to level variations are expected to
have important characteristics \cite{fyodorov94c} which will be analyzed in
future work. However the IPR for the chaotic cavity are confined to a band
around the limiting value of 3, with an average of 3.2. In contrast the IPR
for the disordered cavity, show an increase towards the lower eigenstates,
which tend towards the strong localization regime, showing the dependence of
localization length on the frequency and the transition from regime (a) to
(b). Of course deeper into the diffusion regime in which the localization
length becomes effectively ``infinite'', at large $f$ or small $\lambda $,
these also tend to 3, the value for the chaotic case. The dashed line
represents the functional form for the IPR deduced above from eq.(3), and
which extends up to the regime $\lambda <l$, and which thus represents the
small localization limit of the 1-D $\sigma $-model.

In conclusion, we have been able to study the role of chaos and disorder in
organizing quantum wavefunctions by means of electromagnetic experiments.
The results demonstrate universal properties of chaotic wavefunctions, which
also correspond to disordered systems in the extreme diffusive limit. They
also demonstrate deviations from universality due to localization. The
results appear to be consistent with a proposed notion of a (weaker)
universality class \cite{fyodorov94} for systems with localization, in which
the statistical measures depend on one additional parameter - the
localization length. The disordered billiards used here are the first
experimental realization of such a system in which quantitative results have
been obtained, and the dependence on the parameter $\xi /L$ has been
quantitatively demonstrated, enabling tests of the
nonlinear $\sigma $-model of supersymmetry theory,
originally intended for mesoscopic systems.
These results are of potential relevance to
systems such as mesoscopic devices \cite{baranger94} and acoustic and
electromagnetic waves \cite{mccall91}. 

We thank V.Prigodin, B.Simons, Y.Fyodorov and A.Mirlin for useful
discussions and communicating results prior to publication. This work was
supported by the ONR under Grant No. N00014-92-J-1666.

\begin{figure}
\epsfxsize=8.5cm
\epsffile[20 50 276 240]{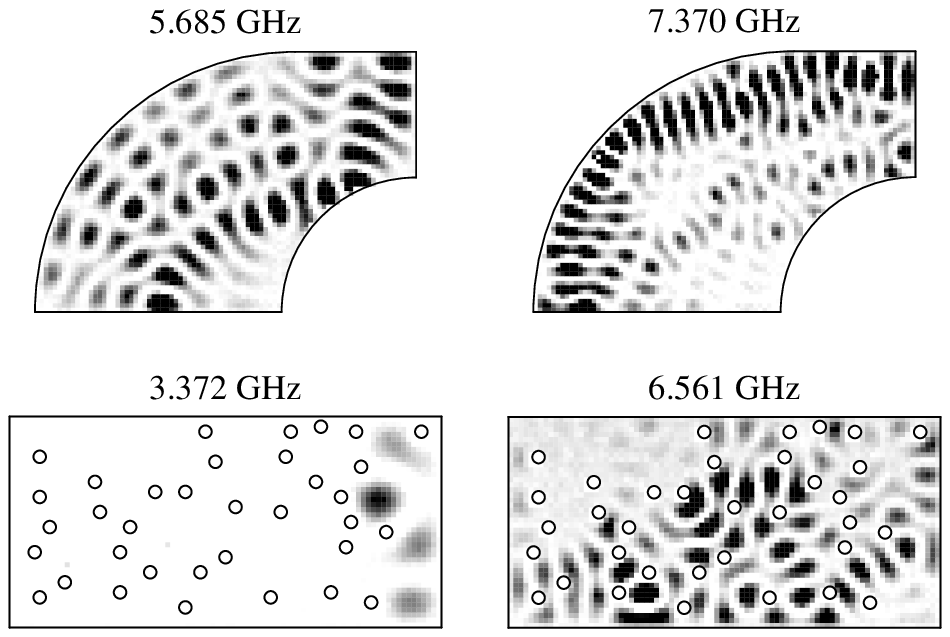}
\caption{Representative eigenfunctions of the chaotic Sinai - Stadium 
billiard (top) and disordered geometry (bottom).}
\end{figure}

\begin{figure}
\epsfxsize=7.5cm
\epsffile[65 65 645 445]{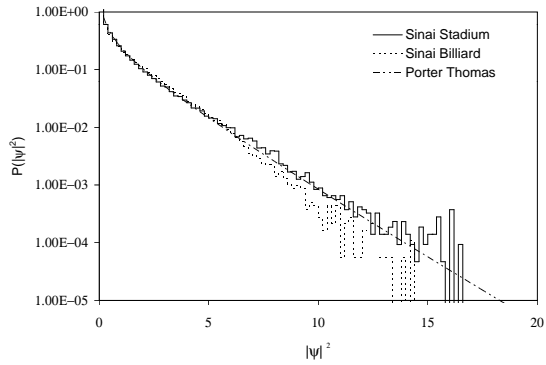}
\caption{Density distribution $P(|{\Psi }|^2)$ for the Sinai-Stadium and the 
Sinai billiard compared with 
the Porter-Thomas form from RMT. While the Sinai-Stadium is in 
excellent agreement, the Sinai Billiard data shows 
slight deviations due to states influenced by bouncing ball orbits.}
\end{figure}

\begin{figure}
\epsfxsize=7.5cm
\epsffile[65 65 645 445]{papfig4.eps}
\caption{Density correlation function $P_2 (r)$ for the chaotic 
geometries. The solid line represents Eq.(2) 
with $c = 2$ as expected from a supersymmetry approach.}
\end{figure}

\begin{figure}
\epsfxsize=7.5cm
\epsffile[65 65 645 445]{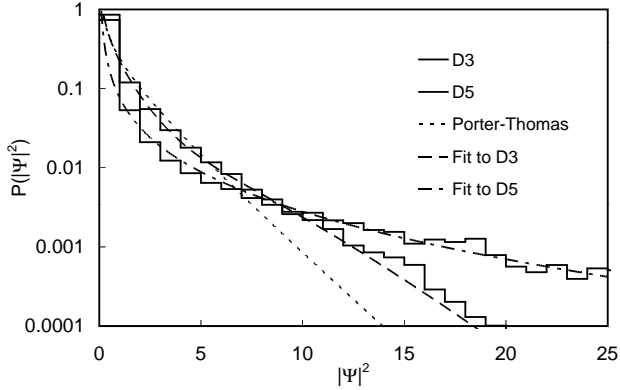}
\caption{$P(|{\Psi }|^2)$ for two realizations of the disordered 
geometries, showing deviations from Porter-Thomas due to 
localization. The dashed line is a fit to Eq. 3. The 
dashed-dotted line is fit to the strong-localization form in the text.}
\end{figure}

\begin{figure} 
\epsfxsize=7.5cm
\epsffile[5 5 645 445]{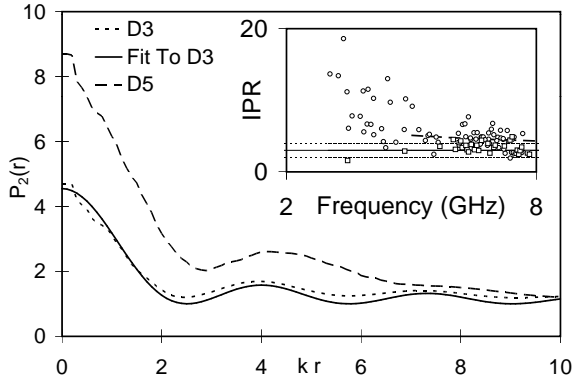}
\caption{$P_2 (r)$ for the disordered wavefunctions 
showing oscillations similar to the chaotic data. The solid line  
represents Eq.(2) with c = 3.5 and kr rescaled by 0.96. (Inset) The inverse 
participation ratio (IPR) vs. frequency for the D3 \& D4 cavities (circles), and
the chaotic cavity (squares). The dotted lines represent the limits of excursion 
of the chaotic data. The dashed curve represents the behavior expected on the
basis of the 1-D $\sigma$ -model.} 
\end{figure}

\begin{tabular}{|c|c|c|c|c|c|}
\hline
     {\bf Cavity}  &  {\bf No of scatterers}  &  {\bf l, cm}  &  {\bf L, cm}  &  {\bf $\xi$/L} &  {\bf c}\\ 
\hline
     D1       &  12  &  9       &  32  &  10  &  2.85\\
     D2       &  27  &  6       &  31  &  5    &  3.14\\
     D3       &  36  &  5       &  31  &  4    &  3.50\\
     D4       &  36  &  5       &  31  &  4    &  3.54\\
     D5       &  72  &  3.5     &  30  &  -    &  -\\ 
\hline
\end{tabular}

\begin{table}
\caption{The various disordered cavities.}
\end{table}

\end{document}